# Adaptation of Black-Box Software Components

**Rolf Andreas Rasenack**

| FH Oldenburg/Ostfriesland/Wilhelmshaven | De Montfort University, |
|---|---|
| Fachbereich Technik, INK | Software Technology Research Laboratory |
| Constantiaplatz 4 | The Gateway |
| 26723 Emden | Leicester LE1 9BH, UK |

rasenack@technik-emden.de

**Abstract:** The globalization of the software market leads to crucial problems for software companies. More competition between software companies arises and leads to the force on companies to develop ever newer software products in ever shortened time interval. Therefore the time to market for software systems is shortened and obviously the product life cycle is shortened too. Thus software companies shortened the time interval for research and development. Due to the fact of competition between software companies software products have to develop low-priced and this leads to a smaller return on investment. A big challenge for software companies is the use of an effective research and development process to have these problems under control.

A way to control these problems can be the reuse of existing software components and adapt those software components to new functionality or accommodate mismatched interfaces. Complete redevelopment of software products is more expensive and time consuming than to develop software components.

The approach introduced here presents novel technique together with a supportive environment that enables developers to cope with the adaptability of black-box software components. A supportive environment will be designed that checks the compatibility of black-box software components with the assistance of their specifications. Generated adapter software components can take over the part of adaptation and advance the functionality. Besides, a pool of software components can be used to compose an application to satisfy customer needs. Certainly this pool of software components consists of black-box software components and adapter software components which can be connected on demand.

## 1. Introduction and Motivation

Today, it is important for software companies to build software systems in a short time-interval, to reduce costs and to have a good market position. Therefore well organized and systematic development approaches are required. Reusing software components, which are well tested, can be a good solution to develop software applications in effective manner. The reuse of software components is less expensive and less time consuming than a development from scratch. But there are important reasons which are addressed and leads to the fact that software components cannot be used directly. One reason is the rapidly change of functional/ non-functional user requirements in real-world software systems. This means the change of functional requirements like data processing, data manipulations, and calculations, as well as the change of non-functional requirements like performance, reliability, and security, impacts directly software applications and must be provided by software components. The other addressed reason is the mismatch between software components. Mismatch appears when the interfaces of corresponding software components are not well designed. Due to the fact of misinterpreting specifications, problems appear in sense of interaction and communication between software components. Both reasons, the change of user requirements and mismatch of software components limit or prohibit their reuse.

Thus software components have to adapt to fulfil these needs. An appropriate approach for adaptation of software components as well as a supporting environment has to be developed. The approach introduced here, presents novel technique together with a supportive environment that enables developers to cope with the adaptability of black box software components.

## 2. Software Components

Software components have some properties and can be characterized by a definition. The term software component is defined in literature in manifold ways. Some definitions try to define the term software component in a general way without technical considerations. Other definitions concentrate on the context in which the software components can be used. For instance software components can be seen as parts of a software system or they can be seen as service provider. To cover all aspects that are related to software components in different context is probably not exhaustive possible. Therefore we concentrates on the most convinced definitions in this topic and excerpt a definition for adaptable software components.

## 2.1 Definitions

As one outcome of the first Workshop on Component-Oriented Programming 1996 (WCOP´96) at European Conference on Object-Oriented Programming 1996 (ECOOP´96) in Linz, Szyperski and Pfister developed the following definition of the term software component:
"*A software component is a unit of composition with contractually specified interfaces and explicit context dependencies only. A software component can be deployed independently and is subject to composition by third parties.*" [Muh97]

In other words this definition describes a software component which consists of combinable pieces software. Pieces of software for instance in the object-oriented programming language Java [**03b] can be a class. This implies that a software component is more coarse-grained than a single class. Logically coherent classes can be compounded to a software component. Well defined interfaces of software components described by a contract are a necessary premise for communicating between software components. A contract between a developer and a client is a precise specification attached to an interface. It covers functional and extra-functional aspects. Functional aspects include the syntax and the semantics of an interface whereas the extra-functional aspects include the quality-of-service guarantees [Szy02].

Additionally software components are designed not only for domain specific applications. They encapsulate its implementation so that it is not possible to have access to the construction details and therefore software components are self-contained. Szyperski abstract this definition into a technical part with considerations such as composition, independence, and contractual interfaces and a market-related part with considerations such as deployment, and third parties [Szy02]. This reflects the practical benefit for the development process of software components.

Another important definition comes from Sametinger. In contrast to the above mentioned definition Sametinger gives a more general definition without consideration of market-related aspects. As one result, in the following definition it is stated that software components are any reusable artefacts. The used term artefact represents different forms of software components. This can be source code or a black-box view that hides the internal details of a software component for instance.

"*Reusable software components are self-contained, clearly identifiable artefacts that describe and/or perform specific functions and have clear interfaces, appropriate documentation and a defined reuse status*". [Sam97]

Self-contained software components means in Sametingers definition that a software component has its own functionality and do not need other software components to provide this functionality. Furthermore software components should be contained in a file and not being spread over many locations then it is identifiable. It has a clear defined interface that hides details that are not needed for reuse. The documentation (specification) must provide enough information to retrieve a software component from a repository, gives information in which context this software component can be used, make adaptations possible. Furthermore the mentioned reuse status of Sametingers definition provides release information of the software component.

The definitions discussed here, include only two representative definitions. But the term coupling between software components are not considered. The necessity to consider the notion coupling, is caused by flexible combining of software components especially for adapt them. In [WY03] the term coupling was taken into account and describes the level of dependencies between interacting software components. Coupling between software components will be differentiated into low coupling or high coupling. The design of highly-coupled software components is based of assumptions between them. Assumptions include for instance every time availability of corresponding software components, syntax for invoking the functionality of interacting software components or data exchange between the software components has to be done every time in the same format. Advantageous of this highly-coupled software components are increasing the performance between the related software components. Disadvantageous is the fact that the software components are specific designed to communicate to each other. This means if requirements are changing for instance in the direction of functionality then all related software components have to adjust to the new situation. But in sense of adaptation of software components it is not acceptable to redesign all related software components because of additional costs, time and may be putting errors in the new developed software. Therefore low coupling is a preferred approach in which software components operate extensive autonomous via interfaces and does not need to be concerned with other software components internal implementation. This is important because changes in one software component have no influence to the corresponding software component. Thus the approach of low coupling is necessary to consider in the definition of software components which can be adapted.

Derived from the above-mentioned discussion, the following combination of definitions will be considered in the area of adaptation of software components:

*A software component is a piece of software which offers a coherent functionality and exhibits certain autonomy by strict encapsulation of the implementation. Flexible combining and separation of software components are achieved by low coupling. Well defined interfaces, responsible for the communication and interaction between components, include a specification which additional describes the behaviour of the software component. The internal structure of a software component will not be considered. Software components can be composed of single software components to achieve an extended functionality.*

## 2.2   Structure of Software Components

The structure of software components characterizes different elements of software components. Yang and Ward define five elements of a software component. That includes code, specification, interface, design and documentation [WY03]. We focus on the approach with the abstract view on three structure elements of a software component. These are:

- Component
- Component interface
- Component specification.

Figure 1 shows a software component with its typical three elements. A well defined *component interface* is required for communication and interaction with other software components. It separates the software component to each other and is described by a corresponding *component specification*. Incoming and outgoing information/ services of software components will be processed by the appropriate provided and required interfaces. The *component* is an element that hides its internal structure for using of third parties. That is, it provides the internal logic (e.g. classes in object oriented programming) which is not present for the client. Hence a component represents certain behaviour and is addressed by the component interface.

157

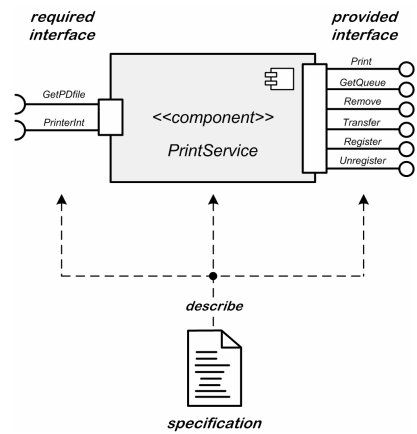

Figure 1: Elements of a Software Component [7]

Software components are represented in different views depending on their abstraction level. The abstraction level defines the different alternatives of the access to the structure of software components. They can be distinguished into black-box, white-box, glass-box and gray-box software components. The scope of research is directed on black-box software components. Black-box software components hide its internal structures from the user and it is not changeable. Of major interest is rather only the behaviour of the black-box software component. The behaviour of this type of software component is characterized by a corresponding specification. It describes the syntax of the component interface and the semantic of the component. A binary form of JavaBeans [**03a] can be a representation of a black-box software component.

## 2.3 Component Mismatch

The increasing productivity of the software development process is attended by the ability of reusable software components to combine (compose) them. Composing applications out of reusable software components leads to rapidly developing in contrast to developing software from scratch. However systematic development of applications from existing software components is an elusive goal. The reason for that is caused by:

- the inability to locate the desired software component and
- the lack of existing software components, as well as
- mismatches between software components to build applications.

To solve the problem of the inability to locate the desired software component it is necessary to provide a component pool that catalogues and categorizes the software components. So that is possible to retrieve a software component for the desired needs. The lack of existing software components leads to development of appropriate new software components. It is obviously that this new components have to store into the component pool.

Reasons that software components cannot interoperate are described by Shaw [Sha95]. To them belong different assumptions about how data is represented, how they are synchronized and what semantics they have.

A promising research topic is the development of novel techniques to detect and cope with mismatches between software components. The knowledge of that technique should be applied to the adaptation process to compose applications to fulfil the expected behaviour of the application.

## 3. Adaptation of Software Components

Component-based software technology represents a software production paradigm that concentrates on the reuse of software components to develop large software systems. The reuse of software components, even so called components-of-the-shelf (COTS), to assemble applications are in practice often problematic. It was hoped that software components can be match together without any change [SH04]. But often in practice the behaviour of a software component is not the same as expected. Due to incompatible interfaces for communication/ interaction between software components and the lack of functionality this problem occurs.

The circumstances that software components cannot be reused ''as-is'' is identified by many researchers. Therefore software components have to be adapted. Approaches for adapting software components exists in manifold directions and can be distinguished depending on their abstraction level into *white-box* or *black-box* approaches. White-box approaches imply the knowledge of internal implementation or structures whereas black-box approaches require knowledge about the components interface and functionality.

### 3.1 Definition of Adaptation

Before discussing the subsequent sections it is necessary to define and classify the term adaptation. It should be clarified for what it is necessary to have adaptation of software components. This means in which direction goes the adaptations process.

In principle *adaptation* of software components is a process to modify them according to changed conditions. Conditions in this context can be identified as the adaptation of software components for their operating in different *environments* or adaptation of software components *behaviour* to assemble applications. The necessity to differentiate the term adaptation into two branches will be supported by two representative definitions:

- definition from Subramanian and Chung for changes in environments:

  *"Adaptation means the change in the system to accommodate change in its environment. More specifically, adaptation of a software system is caused by change from an old environment to a new environment, and results in a new system that ideally meets the needs of its new environment."* [SC02]

- definition from Bosch for component adaptation (behaviour)

  *"The process of changing the component for use in particular application is often referred to as component adaptation."* [Bos99]

Figure 2 depict the identified branches of adaptation and their subdivisions.

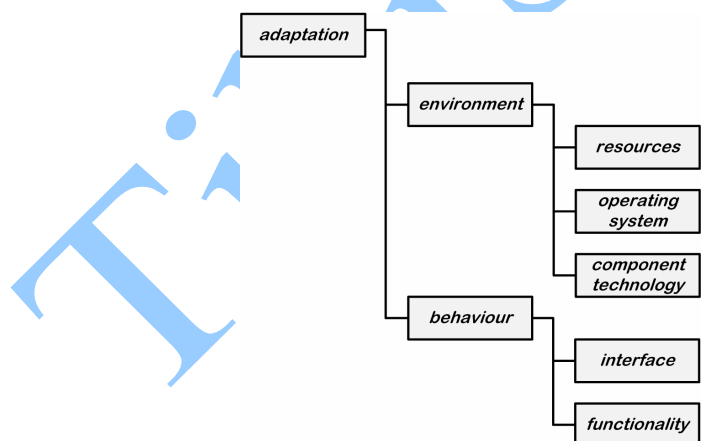

Figure 2: Branches of Adaptation

Generally environments are areas in which software components are applied. We identified three different subbranches of the term environment. The first subbranch is related to the adaptation of software components that are running on different machines. For example software components which are running on a personal computer should be running on a personal digital

160

assistant (PDA). The limited *resources* like memory space or different central processing units (CPU) have to be taken into account for the adaptation process.

Adaptation of software components which are working in one *operating system* and should be working in a different operating system is recognized as second subbranch. As an example for that subbranch filter in the UNIX[1] operating system (e.g. comm filter) are used. Filter in UNIX operating system are software components that gets their data from standard input (keyboard). Afterwards this input data will be process, e.g. compare different input information on common lines, and writes the result to the standard output (terminal). To use that filter for instance in Windows[2] operating system it is necessary to adapt this filter because of different access to the standard input and output operations of both operating systems.

The third subbranch is related to *component technology* in that the components are developed, composed or deployed. Normally software components are composed to operate in a specific environment. For instance Enterprise JavaBeans (EJB) components are expected to be deployed in a suitable EJB server [Nas03]. Using those components in a different way they have to adapt to the target environment.

Obviously the adaptation process regarding to the change in environments is related to have the same functionality of software components. But successful composition of software components will be often prohibited by their behaviour. Caused by interconnection problems and/ or the lack of functionality this difficulties occur. This mainly is due to the fact that the development process of software components uses different standards respectively standards are not consistently in use. Apparently software components were developed and pass tests without any claim. But problems occur while connecting them together. Not compatible *interface* make sure that the software components cannot interacting. On the other hand the lack of *functionality* makes sure that the resulting applications do not achieve the user requirements.

We identified that there is a promising potential to research in the field of adaptation the behaviour of software components. Many researchers focus on adaptation software components for changing environments. But there is a lack of research activities to accommodate software components respective their behaviour. Based on semantics of software components, a development environment simplifies developers work. The environment contains a management part, a data base as well as an analyser part for generating adapter software components.

---

[1] UNIX is a trademark of The Open Group since 2007, an industry standards consortium
[2] Windows is a trademark of Microsoft Corporation

## 4. Proposed Adaptation Framework

The architectural design of a framework for adapted software components provides an environment in which the black-box software components are checked whether they can match or not. The principle of that check is based on comparing the semantics of the component specification with the project specification to adapt software components. Additional this framework can be used to extend the functionality of an application by inserting software components. This chapter describes in general the approach to cope with that components and show how the framework engage into the application. Obviously we can recognize a framework part and the application part of the adaptation framework. The framework part provides the necessary tools and algorithms to process black-box software components. The application part defines the input and output communication to the framework part and can be seen as developers working environment. The framework part does its work in the background. The application part shows a simple scenario of a project which consists of two software components (Figure 3). Software component A and software component B should be communicate each other, but they cannot match because of different interfaces. An adapter software component is generated by the *Analyser* and accommodates the interfaces. Chapter 4.1 describes this process more detailed.

## 4.1 The Adaptation Framework

It is assumed that a developer build up an application with existing software components. This application is stored as a software project and is described by a project specification. That specification describes in which context the application will be used. The example application consists of two software components A and B which provide an appropriate functionality and are described by a component specification. Obviously we have to distinguish two types of specifications. One is called *project specification* and represents the target state of the application. The second type is called *component specification* and represents the as-is state. During the development process it was visible, that the software components can not match together. At this situation the framework engage into the application.

Taking into consideration the project specification, the component specifications will be read (1) and afterward compared to determine (2) the demand of functionality of a sought software component or software adapter. In Figure 3 this is depict by *Analyser*. The administration between *Component Pool* and *Analyser* will be performed by the *Linkage Management*. The *Analyser* informs the *Linkage Management* about the demand of necessary functionality for the application. Thereupon the *Linkage Management* makes a request (3) to the *Component Pool* whether there is an appropriate software component or software adapter stored in the data base. After searching the data base it returns (4) the result. Different results are possible. One result can be that there is a software component or software adapter for the adaptation process. Another result can be that there is no component or adapter, so it is necessary to develop a new software component or software adapter to fulfil the needs of the application. In this example we consider that there is no component or adapter. Therefore the *Linkage Management* invites (5) the *Analyser* to take the information of the comparison process to generate (6) an adapter software component. The generated software component provides the necessary functionality and interfaces for communicating and interacting between software component A and software component B. Now the *Linkage Management* integrate (7) the generated software adapter into the application. To extend the *Component Pool* the new adapter software component is stored. After adaptation the analysing procedure can be applied again for verification of the application.

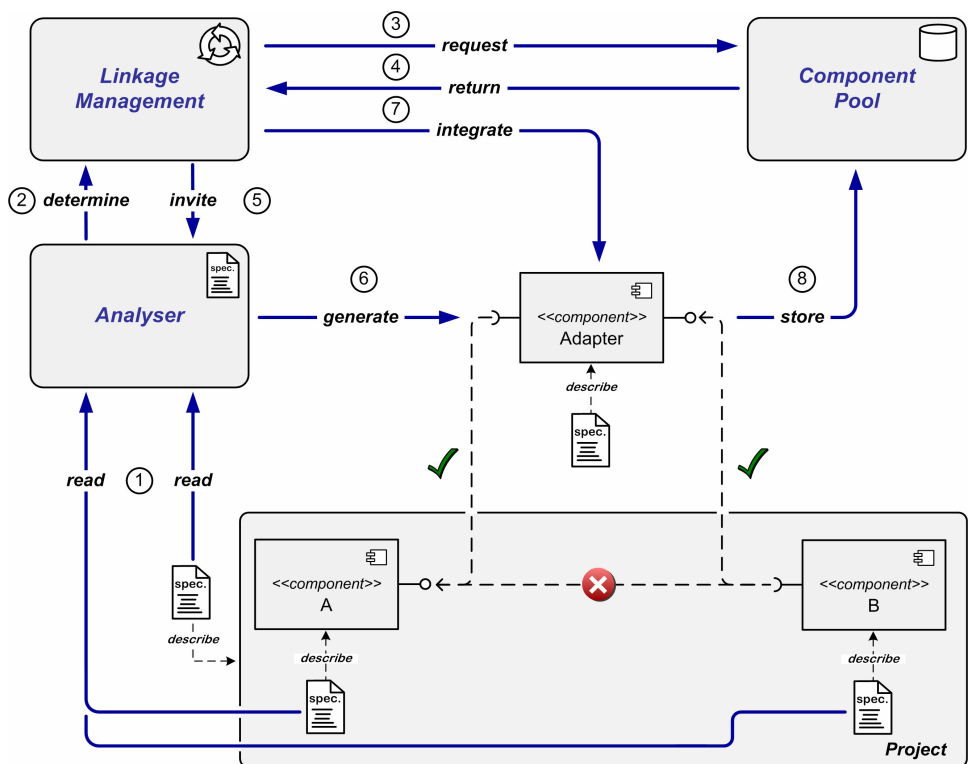

Figure 3: Adaptation Framework

## 4.2 The Analyser

The Analyser is an important part of the adaptation framework. It provides the process for analysing specifications and generating an appropriate software adapter. Figure 4 depicts the systematic generation of software adapter. What kind of infrastructure the *Analyser* uses is described subsequent.

A challenge is to describe semantic and to analyse compatibility for adaptation of black-box software components. To describe the architecture for manipulation with black-box software components one possible solution can be the use of architecture description languages (ADL) for example like the well known unified modeling language (UML). In addition the behaviour of software can be specified in an abstract way with the use of standard languages like Object Constraint Language (OCL), Semantic Web or Ontology's. Another solution to solve the problem of describing and processing the behaviour of black-box software components is the use of Meta Information Definition Language (MIDL).

This generic language is developed and described by Wolke [Wol07]. With MIDL it is possible to specify the semantic of software components. It can be seen as a macro language that can individually accommodate to specific needs. In this case meta information will be specified that can be processed by special processors to generate adapter software components.

The processor for generating adapter software components is called Meta Information Processing Tool (MIPT) and is shown in Figure 4. With the assistance of an Abstract Syntax Language Tree (ASLT) the processing (analysis) shall be conducted.

The ASLT is the representation of object-oriented structures (packages, classes, variables and methods), which become visible as hierarchical elements (nodes). The ASLT is the basis for variants of implementation and/or views (UML class diagram, Nassi Sheidermann diagram etc.), which are made available to the developer. Each view offers to the developer a special sight of a project.

Thereby only certain parts of a project will be represented, the remaining other parts becoming invisible by folding. The ASLT is the model for the administration of hierarchic elements and particularly for the representation and/or finding of meta information, which is intended for semantic check of software components [WYSR04, YWSR04].

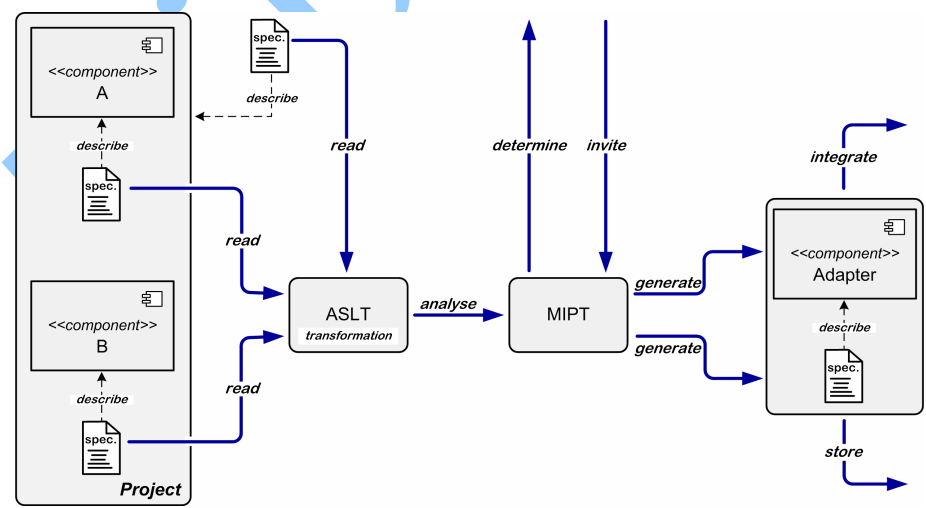

Figure 4: Analysing process

165

Meta information covers extended properties; they are attached as nodes within the ASLT. Meta information can be processed by software tools (MIPT) [Wol06]. Meta information within the ASLT permits the analysis and manipulation with programmed logic like software components.

## Conclusion

Due to the use of new technologies, error correction (e.g. mismatched interfaces) and implementation of newer functionalities for example the fulfilment of user requirements for software components, necessarily one or more software components of an application has to be adapted or software components must be added. The described framework assists the developers work by managing and generating software components respective software adapter.

The framework engages into the developed application (software project) by inserting software components or adapter. The framework for adaptation compares different specifications that describes semantic of software components. The specifications can be seen as meta information.

For that purpose the specifications respective meta information are transferred into a hierarchical structure (ASLT) because the comparison algorithm can be simply applied in an ASLT. Afterward a processor (MIPT) generates a software adapter or a software component that fulfils the same requirements will be retrieve from the *Component Pool*.

The software development process will be more transparent because of a comparison algorithm that makes sure that the right software adapter will be generated or the right software component will be retrieve from the *Component Pool*. The development time is reduced and inconsistencies are avoided by check of interfaces. At this concept it is favourable that the proving of software components is based on semantic and not on test cases.